\documentclass{l4dc2027}

\title[Learning to Clear Peer-to-Peer Energy Markets]{Learning to Clear Peer-to-Peer Energy Markets}
\usepackage{times}
\usepackage{appendix}
\usepackage{graphicx}
\usepackage{float}
\usepackage{booktabs}
\newtheorem{assumption}{Assumption}
\usepackage{booktabs}
\usepackage{multirow}
\usepackage{amsmath}
\usepackage{amsfonts}
\usepackage{xcolor} 
\usepackage{float}
\usepackage{algorithm, algorithmic}
\usepackage{balance}
\usepackage{hyperref}

\coltauthor{\Name{Lisa {Innocenti Uccini}} \Email{lisaiu@kth.se}\\
 \Name{Andrea Martin} \Email{andrmar@kth.se}\\
 \Name{Giuseppe Belgioioso} \Email{giubel@kth.se}\\
 \addr Department of Decision and Control Systems, KTH Royal Institute of Technology, Stockholm, Sweden
 }

\begin{document}

\maketitle

\begin{abstract}%
Peer-to-peer energy markets are a promising paradigm for integrating distributed energy resources. A large body of work models their clearing as distributed optimization or generalized Nash equilibrium (GNE) problems which let prosumers coordinate trades and local resources without a central operator. Yet, the computational and communication burden of distributed optimization and GNE-seeking algorithms poses a major challenge to deployment at scale. We propose a learning-to-optimize framework for decentralized markets that uses historical data to accelerate clearing algorithms without additional communication. To enforce convergence by design, we train summable augmentations to a baseline algorithm, using graph neural networks to learn permutation-equivariant updates for markets of varying size and topology. We instantiate the framework for computing a variational GNE in the market model of \cite{cadre_peer-peer_2020} with a distributed Douglas–Rachford algorithm, but the idea extends to other models and base algorithms. Numerical experiments show improved convergence and illustrate the robustness of our method to out-of-distribution instances.

\end{abstract}


\section{Introduction}

The ongoing transition of power systems, driven by the large-scale deployment of distributed energy resources at the grid edge, is reshaping the role of end users. Consumers are increasingly becoming prosumers, namely, agents that not only use energy, but also generate it and actively participate in the system operation~\citep{parag2016electricity, tushar2020peer}. As a result, conventional top-down energy markets are becoming less suited to these new forms of participation, motivating decentralized paradigms~\citep{pinson2023may}. Peer-to-peer (P2P) energy markets, which enable prosumers to exchange energy directly with one another over the distribution network~\citep{liu2019peer}, have emerged as a promising framework for this transition, as reflected by a growing number of pilot projects and demonstrations~\citep{zhang2017review, ableitner2019quartierstrom, mengelkamp2018designing}.\looseness-1

Two main approaches have been used to model how energy exchanges are determined in P2P markets~\citep{sousa2019peer,tushar2018transforming,islam2024review}. Optimization-based formulations cast market clearing as a single welfare-maximization problem, solved either centrally by a system operator~\citep{cadre_peer-peer_2020} or by decomposing the welfare objective across prosumers via consensus schemes~\citep{sorin2018consensus, baroche2019prosumer, moret2020loss}. To account for strategic interactions between self-interested agents, game-theoretic formulations instead model market clearing as a generalized Nash equilibrium problem (GNEP), where every prosumer simultaneously minimizes its own local cost subject to individual and shared trading and network constraints~\citep{belgioioso2022operationally, cadre_peer-peer_2020}. The resulting market outcome is characterized by a GNE, a stable market configuration in which no prosumer can improve their local cost via a unilateral deviation that remains feasible given the decisions of all other prosumers~\citep{facchinei2010generalized}.\looseness-1

A key challenge to the deployment of P2P energy market structures in real-world applications is the computation and communication complexity of clearing the market when %
many of self-interested prosumers engage in P2P transactions~\citep{moret2018negotiation, pinson2023may}. Building on connections between GNEs and variational inequality theory to reformulate GNEPs as structured inclusion problems~\citep{facchinei2003finite}, several distributed GNE seeking algorithms have recently been proposed~\citep{liu2017energy, belgioioso2022operationally, cadre_peer-peer_2020}. %
These iterative schemes guarantee asymptotic convergence to a market equilibrium while operating under the information constraints inherent to P2P energy markets, where prosumers update their decisions based on local information and messages exchanged with neighboring agents only. However, the large number of iterations required by these algorithms to converge, together with the communication cost of information exchanges over the trading network, represents major threats to the practical deployment and the robust operation of autonomous P2P energy markets~\citep{moret2018negotiation, baroche2019prosumer}.

Learning to optimize (L2O) has emerged as a promising paradigm for improving the efficiency of iterative optimization algorithms and tailoring their update rules to the structure and statistics of specific problem classes, as characterized by representative training instances~\citep{chen2022learning}.\footnote{This stands in stark contrast to classical acceleration schemes for fixed-point algorithms, such as Anderson acceleration~\citep{anderson1965iterative} or SuperMann~\citep{themelis2019supermann}, that combine hand-crafted momentum terms, safeguard mechanisms, and line searches with global residual evaluation to improve worst-case convergence rates.} However, while the flexibility of neural network (NN) architectures allows learning efficient algorithms from data~\citep{andrychowicz2016learning}, directly replacing hand-crafted algorithmic updates with learned components may compromise their convergence guarantees, a critical limitation in P2P energy markets, where convergence is key to stable market outcomes. To address this issue, model-based L2O approaches restrict learning to warm-starting classical algorithms~\citep{sambharya2024learning} or tuning hyperparameters~\citep{ichnowski2021accelerating, doerks2025learning}, without changing the structure of their update rules and thus offering limited control over their transient performance. To fully exploit the expressiveness of NN updates,~\citet{heaton2023safeguarded, premont2022simple} design safeguarding mechanisms that conditionally switch between classical and learned updates, whereas~\citet{martin2024learning, banert2024accelerated, martin2025learning, martin_learning_2026} parametrize classes of convergent algorithms by characterizing perturbations to the update rules of classical optimizers that preserve convergence by design. Despite these advances, the application of L2O to distributed P2P market clearing remains challenging, as learned updates must ensure convergence while complying with the decentralized information structure of the underlying GNEP, avoiding the global coordination often required by safeguarding mechanisms.

\emph{Contributions:} 
In this paper, we exploit historical data to accelerate distributed market clearing, towards enabling P2P energy markets at scale, without requiring additional communication and while preserving convergence guarantees.
Specifically, instead of using NNs to predict equilibria directly~\citep{bichler2021learning, wang2020multi, jungel2025wardropnet, bemporad2026learning, mckenzie2024three},
we introduce a parametrization of learned augmentations to a  splitting method that guarantees their summability by construction, and we prove convergence by analyzing the resulting algorithm through the lens of inexact fixed-point iterations~\citep{martin_learning_2026}.
P2P energy markets must be cleared at every trading interval, e.g., 
every 15 or 30 minutes in existing pilot projects~\citep{mengelkamp2018designing, ableitner2019quartierstrom}, so that historical data naturally accumulate over time and provide representative samples to be cleared in the future.
We further show how graph neural networks (GNNs)~\citep{zhou2020graph} can be used to parametrize these augmentations, yielding topology-aware and permutation-equivariant learned updates that preserve the local information structure of distributed GNE seeking algorithms and can accommodate different market configurations. Last, we evaluate our approach on synthetic instances of the P2P market model of~\citet{cadre_peer-peer_2020} to demonstrate its potential for accelerating convergence and reducing communication overhead while maintaining robustness to out-of-distribution instances. The code is available at %
\href{https://github.com/zero-lab-kth/learning-to-clear-P2P.git}{\texttt{github.com/zero-lab-kth/learning-to-clear-P2P.git}}.

\emph{Notation:} 
When a finite dataset is considered, $i$ indexes samples, e.g., $x_n^{k,i}$ denotes the local variable of agent $n$ at iteration $k$ for sample $i$, sample indices are omitted when referring to a generic instance. When relevant, the dependence of a quantity on trainable parameters is made explicit, e.g., $x^{k,i}(w)$, and is otherwise omitted for simplicity. A superscript $^\star$ denotes an equilibrium or an optimizer, according to context. The operator $\operatorname{col}((x_n)_{n\in\mathcal N})$ denotes the column-wise stacking of the vectors $x_n$. 
For a set $\mathcal S$, $\operatorname{dist}(x,\mathcal S)$ denotes the Euclidean distance of $x$ from $\mathcal S$. Given a nonempty closed convex set $\mathcal X\subseteq\mathbb R^d$ and a mapping $F:\mathbb R^d\to\mathbb R^d$, the solution set of the variational inequality $\operatorname{VI}(\mathcal X,F)$ is
$\operatorname{SOL}(\mathcal X,F):=\{x^\star\in\mathcal X\mid\langle F(x^\star),x-x^\star\rangle\geq 0,\ \forall x\in\mathcal X\}$.

\section{Problem Formulation}
\label{sec:probformulation}
In this section, we formalize the problem of designing an efficient algorithm for clearing P2P energy markets. Following~\citet{cadre_peer-peer_2020}, we first model prosumer interactions as generalized noncooperative games with reciprocity constraints. Then, we state key convergence and implementability requirements that underpin the design of distributed market clearing algorithms.
\subsection{Game-theoretic model of P2P energy market}
We consider a P2P energy market whose trading network is modeled as a connected undirected graph $\mathcal{G} = (\mathcal{N}, \mathcal{E})$, where
$\mathcal{N} = \{0, \ldots, N{-}1\}$ is a set of $N \in \mathbb{N}$~nodes, each corresponding to a different prosumer,
and $\mathcal{E} \subseteq \mathcal{N} \times \mathcal{N}$
encodes peer-to-peer trading links.
We denote by $\Omega_n \subseteq \mathcal{N}$ the set of neighbors of $n\in\mathcal{N}$, namely, the set of prosumers with whom $n$ can directly trade.%

Each prosumer~$n$ controls three types of decision variables: the flexible demand $D_n \geq 0$, which represents the energy consumed to achieve a maximum comfort level by meeting a  pre-specified target demand $D_n^{\text{ref}}$; the controllable local generation $G_n \geq 0$ (also known as flexibility activation), consisting of dispatchable resources such as micro combined-heat-and-power (CHP) units or storage facilities that the agent can actively modulate; and the bilateral trades $q_{mn} \in \mathbb{R}$, where, for each $m \in \Omega_n$,
$q_{mn} > 0$ means that $n$ buys from~$m$ and $q_{mn} < 0$ means that $n$
sells to~$m$. Each prosumer~$n$ is also equipped with a stochastic renewable energy source (RES) generation
$\Delta G_n \geq 0$ (e.g., photovoltaic output), treated as a
fixed parameter not subject to optimization.
Overall, the local decision vector of prosumer $n \in \mathcal{N}$ is
$x_n = (D_n,\, G_n,\, q_n^\top)^\top \in \mathbb{R}^{2+|\Omega_n|}$,
where $q_n = (q_{mn})_{m \in \Omega_n}$. We let $x=(x_n)_{n\in\mathcal{N}} \in \mathbb{R}^{n_x}$ collect all prosumers' decisions, with $n_x:=\sum_{n\in\mathcal{N}}{(2+|\Omega_n|)}$.

We define the local feasibility set of prosumer~$n \in \mathcal{N}$ by collecting all operational and energy balance constraints that only involve its own decision variables, namely,
\begin{equation}\label{eq:local_set}
  \mathcal{X}_n^{\mathrm{\text{loc}}}(\theta_n) \!=\!
  \left\{\! x_n :
  \begin{array}{l}
    D_n\in[\underline{D}_n, \overline{D}_n],\;
    G_n\in[\underline{G}_n, \overline{G}_n],\; 
    D_n = G_n + \Delta G_n + Q_n,\\[2pt]
    q_{mn} \leq \kappa_{mn},\;\forall m \in \Omega_n, 
  \end{array}
  \!\right\}\!,
\end{equation}
where $Q_n=\sum_{m\in\Omega_n} q_{mn}$ is the net import at node $n$ and $\kappa_{mn}=\kappa_{nm}>0$ is the maximum interconnection capacity. The parameter vector $\theta_n\in\Theta_n\subseteq\mathbb{R}^{p_n}$ in~\eqref{eq:local_set}, which $p_n$ depends on $\Omega_n$, collects data relevant to prosumer $n$, i.e., the RES output, the target demand, and the local objective
parameters in~\eqref{eq:cost}, so that the market is parametric in $\theta = (\theta_n)_{n\in\mathcal{N}} \in \Theta := \prod_{n\in\mathcal{N}} \Theta_n \subseteq \mathbb{R}^{p}$, where $p := \sum_{n\in\mathcal{N}} p_n$. Accordingly, we define a P2P market-clearing instance as a pair $\eta = (\mathcal{G}, \theta)$, with $\mathcal{G}$ a trading graph and $\theta$ a parameter vector of dimension compatible with $\mathcal{G}$. Since each $q_{mn}$ is
bounded above by $\kappa_{mn}$ and, through the energy balance, from below, $\mathcal X^{\mathrm{loc}}_n(\theta_n)$ is a compact polyhedron. Let $\mathcal{X}^{\text{loc}} (\theta)=\prod_{n\in\mathcal{N}} \mathcal{X}^{\text{loc}}_n(\theta_n)$ be the local feasibility set of the market. 
Trading prosumers are coupled through the trade reciprocity condition:
\begin{equation}\label{eq:reciprocity}
  q_{mn} + q_{nm} \leq 0,
  \qquad \forall\, \{n,m\} \in \mathcal{E}.
  \vspace{-0.2cm}
\end{equation}
Modeling reciprocity as an inequality rather than an equality allows the energy offered to be only partially absorbed. We collect the coupling constraints in the reciprocity set $\mathcal{X}^{\mathrm{rec}} =
 \{x \in\mathbb{R}^{n_x}: \eqref{eq:reciprocity} \text{ holds}\}$, a polyhedral cone obtained as the
intersection of $|\mathcal{E}|$ half-spaces, each involving a distinct pair of
trade variables.
Each prosumer~$n$ aims to minimize a local cost function
\begin{equation}\label{eq:cost}
J_n(x_n; \theta_n) = \underbrace{\tfrac{1}{2} a_n G_n^2 + b_n G_n + d_n}_{C_n(G_n, \theta_n)} + \underbrace{{\textstyle\sum_{m \in \Omega_n}} c_{nm}\, q_{mn}}_{\tilde{C}_n(q_n; \theta_n)} - \underbrace{(\tilde{b}_n-\tilde{a}_n (D_n - D_n^{\text{ref}})^2)}_{U_n(D_n; \theta_n)},
\end{equation}
whose three components are the convex quadratic generation cost $C_n$, the trading cost $\tilde C_n$, whose coefficients $c_{nm}>0$ encode per-unit trading preferences, e.g., reflecting proximity or policy-induced charges, and the usage comfort $U_n$, strictly concave and maximized at $D_n^{\text{ref}}$, capturing a preference for a target consumption level. All parameters $a_n, b_n, d_n, \tilde{a}_n, \tilde{b}_n$ are positive.
Given a specific P2P market-clearing instance $\eta$, each prosumer~$n$ faces the following parametric optimization problem:

\begin{equation}\label{eq:game}
\begin{aligned}
   \min_{x_n}\quad  J_n(x_n; \theta_n) \quad
   \text{s.t.}\quad 
    x_n \in \mathcal{X}_n^{\mathrm{\text{loc}}}(\theta_n), \quad (x_n,x_{-n}) \in \mathcal{X}^{\mathrm{rec}}.
\end{aligned}
\end{equation}

Note that prosumers' decisions on local operational and trading profiles are coupled through the reciprocity constraints in $\mathcal{X}^{\mathrm{rec}}$, making the collection of problems \eqref{eq:game}, one for each prosumer $n\in \mathcal{N}$, a generalized noncooperative game \citep{facchinei2010generalized}. Solutions to such a game are known as generalized Nash equilibria (GNEs), as defined next.
\begin{definition}[GNE]\label{def:gne}
For every $\theta \in \Theta$, let $\mathcal{X}(\theta):= \mathcal{X}^{\mathrm{loc}}(\theta) \cap \mathcal{X}^{\mathrm{rec}}$. A tuple of feasible strategies $x^\star = (x^\star_0, \ldots,x^\star_{N-1}) \in \mathcal{X}(\theta)$ is a GNE of the game~\eqref{eq:game} if $x^\star_n$ solves \eqref{eq:game} for all $n \in \mathcal{N}$.
{\hfill $\square$}
\end{definition}

Such $x^\star$ represents a stable market outcome at which no prosumer can unilaterally reduce their cost by deviating to a feasible strategy.
Among the possibly many GNEs, we focus on variational GNEs (v-GNEs), a refinement with uniform marginal price across each trade.

\begin{definition}[v-GNE]\label{def:vgne}
Let $F(x;\theta):= \operatorname{col}\big(\nabla_{x_n} J_n(x_n;\theta_n)\big)_{n \in \mathcal{N}}$
denote the pseudo-gradient of the game~\eqref{eq:game}. A v-GNE is a
tuple of feasible strategies $x^\star \in \mathcal{X}(\theta)$ solving the 
$\mathrm{VI}(\mathcal{X}(\theta), F(\cdot;\theta))$. 
{\hfill $\square$}
\end{definition}

\subsection{Distributed market-clearing design objectives}
\label{sec:2b}
Equilibria of~\eqref{eq:game} are computed iteratively through distributed algorithms, in which each prosumer updates its operational profile based on local data and messages exchanged~\citep{baroche2019prosumer, belgioioso2022operationally, cadre_peer-peer_2020}.
These algorithms can generally be described by a collection of local update maps $\pi=\{\pi_n^k\}_{n\in\mathcal{N}, k\geq0}$ executed by prosumers, where, at iteration $k$, agent $n$ updates their local variables according to
\begin{equation}\label{eq:recursionN}
  x_n^{k+1} = \pi^k_n\Big(\{x_n^t\}_{0\leq t\leq k},\, \theta_n,\,
  \{q_{nm}^k\}_{m\in\Omega_n}\Big),
\end{equation}
and $q_{nm}^k$ denotes the trade information received from neighbors.
We consider the class $\Pi$ of structurally admissible schemes whose local update mappings support arbitrary neighborhood sizes and are permutation invariant with respect to neighbor ordering. Such schemes can be applied to markets with varying numbers of prosumers and trading topologies.

Let $\mathbb{P}$ denote the distribution over market-clearing instances, of which historical data provide samples. We formulate the update-rule design as an optimization problem of the form
\begin{subequations}
\label{eq:optimal-iteration-design-problem}
     \begin{align}
         \label{eq:meta_loss}
          \inf_{\pi \in \Pi} &~~ \mathbb{E}_{\eta \sim \mathbb{P}} \left[\mathcal{L}_K(\{x^k\}_{k=0}^{K}, \eta)\right]\\
         \text{s.t.} &~~ \eqref{eq:recursionN},
         \quad \forall n\in\mathcal N,\ \forall k\geq 0, \label{eq:meta_loss_constr1}\\
         &~~ \lim_{k \to \infty}\operatorname{dist}(x^k, \mathrm{SOL}\big(\mathcal{X}(\theta),   F(\cdot;\theta)\big)) = 0, \quad \forall x^0 \in \mathbb{R}^{n_x}\label{eq:meta_loss_constr2},%
         \end{align}
 \end{subequations}
where $ \mathrm{SOL}\big(\mathcal{X}(\theta),
  F(\cdot;\theta)\big)
$ denotes the set of all v-GNEs of \eqref{eq:game}. In \eqref{eq:optimal-iteration-design-problem}, the constraint $\pi\in\Pi$ encodes the distributed and structural requirements introduced above,~\eqref{eq:meta_loss_constr2} enforces convergence to a v-GNE for every admissible market instance and initialization, whereas ~\eqref{eq:meta_loss} instead favors improved average-case performance, according to a user-defined performance criterion $\mathcal{L}_K$, over instances $\eta \sim \mathbb{P}$. For instance, to promote rapid convergence of the iterates $x^k$ to the v-GNE set, one could select the algorithmic cost function
\begin{equation}
\label{eq:ideal_loss}
    \mathcal{L}_K
    \left(\{x^k\}_{k=0}^{K},\eta\right)
    :=
    \sum_{k=1}^{K}
    \alpha_k\,
    \operatorname{dist}
    \left(x^k,\mathrm{SOL}\big(\mathcal{X}(\theta),   F(\cdot;\theta)\big)\right),
\end{equation}
where $\alpha_k>0$ are iteration weights. However, as the set of v-GNEs $\mathrm{SOL}\big(\mathcal{X}(\theta), F(\cdot;\theta)\big)$ is generally not available in P2P energy market applications, in Appendix~\ref{app:gnn1} we replace \eqref{eq:ideal_loss} with a proxy based on the empirical fixed-point residual to enable self-supervised training.
\section{Learning-Enhanced Market Clearing}
\label{sec:learned_dr}
We develop a market-clearing algorithm that leverages historical market data to improve finite-time performance, while retaining distributed implementability, applicability across heterogeneous markets, and convergence to a v-GNE, as required by the design problem in Section~\ref{sec:2b}.

Rather than searching over arbitrary admissible update rules $\pi\in\Pi$, we consider a structured class obtained by augmenting a fixed distributed equilibrium-seeking algorithm with a learned component aimed at improving its transient behavior~\citep{martin_learning_2026}.
Specifically, we use a Douglas--Rachford (DR) splitting method as the baseline solver. The learned component enters as an additive perturbation to the iteration, constructed to satisfy the distributed, structural, and convergence requirements of the market clearing, while retaining the baseline fixed-point structure.

\subsection{Learning-enhanced Douglas--Rachford algorithm} \label{sec:3a}

Given a market-clearing instance $\eta = (\mathcal{G}, \theta)$, the proposed scheme is summarized in Algorithm~\ref{alg:learned_dr} below, executed in parallel by every node $n\in\mathcal{N}$.
Lines~3-7 implement a fully-distributed relaxed DR iteration to solve the VI in Definition \ref{def:vgne}. The learned term $v_n^k$ (line~8) is evaluated locally by each prosumer using a distributed GNN-parametrized map trained offline on instances drawn from $\mathbb{P}$; its architecture, internal state, and trainable parameters are specified next.
Note that the sequence \(\{x^k\}_{k\geq 0}\) is the auxiliary fixed-point sequence of the perturbed DR splitting and does not, in general, represent a feasible market allocation.  The market-clearing solution is instead recovered as the limit of the shadow
iterates \(\{ \tilde x^k\}_{k\geq 0} \), where $\tilde x^k=\operatorname{proj}_{\mathcal{X}^{\mathrm{rec}}}(x^k)$ satisfies trading reciprocity at all iterations $k$.
A formal derivation is provided in Appendix \ref{app:proof}.

\begin{algorithm}[H]
\caption{Learning-enhanced Decentralized Market Clearing}
\label{alg:learned_dr}
\begin{algorithmic}[1]

\STATE \textbf{Input:} $\theta_n \in \Theta_n$;
       \textbf{Parameters:} $\lambda\in(0,2)$, $\gamma>0$, $\tau>0$, $\rho\in(0,1)$;
       \textbf{Init:} $x_n^0=(D_n^0,G_n^0,q_n^0)$
\FOR{$k = 0, 1, 2, \ldots$ until convergence}
  \STATE Send $q_{mn}^{k}$ to (receive $q_{nm}^{k}$ from) each $m \in \Omega_n$
  
    \STATE
    $\tilde x_n^k = (D_n^k,G_n^k,\tilde q_n^k)$,
    \quad
    $\tilde q_{mn}^{k} =
    q_{mn}^{k} -
    \tfrac{1}{2}\max\{0,q_{mn}^{k}+q_{nm}^{k}\}$,
    $\ \forall m\in\Omega_n$
        
  \STATE $r_{mn}^{k} = 2\tilde q_{mn}^{k} - q_{mn}^{k}, \quad \forall m \in \Omega_n$
  \STATE
    $
    z_n^k
    =
    \underset{\xi\in\mathcal{X}_n^{\mathrm{\text{loc}}}(\theta_n)}
    {\operatorname{arg\,min}}
    \left\{
        \gamma J_n(\xi;\theta_n)
        +
        \frac{1}{2}
        \left\|
            \xi
            -
            (D_n^k,G_n^k,r_n^k)
        \right\|^2
    \right\}
    $

  \STATE $\hat x_{n}^{k} = x_{n}^{k} + \lambda\,(z_n^{k} - \tilde x_n^k)$
  \STATE $v_n^k = \rho^k\text{GNN}(x^k_n, \theta_n, \{q^k_{nm}\}_{m\in\Omega_n}, \text{local state})$
 \hfill {\footnotesize$\triangleright$ Sec.~\ref{sec:gnn}}

  \STATE $x_n^{k+1} = \hat x_n^{k} + v_n^k$
        
\ENDFOR
\RETURN $\tilde x^{k+1}$%
\end{algorithmic}
\end{algorithm}

\begin{remark}[baseline iteration]
The DR splitting has several favorable properties in this context: it works under mere monotonicity of the pseudo-gradient and its resolvent steps decompose according to trading network topology of the market. In principle, however, any other distributed market-clearing algorithm with a fixed-point interpretation, e.g.,  \citep{sorin2018consensus, baroche2019prosumer, moret2020loss, belgioioso2022operationally, cadre_peer-peer_2020}, can be used as a baseline.
{\hfill $\square$}
\end{remark}

\subsection{Distributed graph neural network parametrization}
\label{sec:gnn}

The learned component in Algorithm~\ref{alg:learned_dr} must preserve the distributed and structural requirements defined by $\Pi$. 
To this end, we adopt a message-passing GNN~\citep{qian2024exploring}\footnote{Standard recurrent architectures commonly used in L2O do not directly satisfy these requirements. For example, the LSTM-based parametrization in~\citet{martin_learning_2026} operates on the joint iterate $x^k$, requiring centralized information collection, and its input and output dimensions scale with the problem size.} whose trainable weights are shared across nodes and trading links and parametrize local update rules~\eqref{eq:recursionN} of the form
\begin{equation}
  x_n^{k+1} = \pi_{\mathrm{DR},n}^k\big(x_n^k,\theta_n,\{q_{nm}^k\}_{m\in\Omega_n}\big) + v_n^k,
  \qquad
  v_n^k = \rho^k u_n^k,\quad \rho\in(0,1),
  \label{eq:decaying_perturbation}
\end{equation}
where $\pi_{\mathrm{DR},n}^k$ denotes the local DR update and $\rho$ is a fixed parameter controlling the time scale over which the GNN output $u_n^k$ remains active. We proceed to define how each agent can generate $u_n^k$ only using local information and variables already exchanged in line~3 of Algorithm~\ref{alg:learned_dr} to construct the GNN messages, therefore avoiding additional communication.

At iteration $k$, prosumer $n$ constructs a node feature $h_n^k\in\mathbb{R}^{d_v}$, describing its local state, and, for each $m\in\Omega_n$, an edge feature $e_{mn}^k\in\mathbb{R}^{d_e}$, describing the corresponding bilateral trade as
\begin{equation}
\begin{aligned}
  h_n^k &= \big(\hat D_n^k,\, \hat G_n^k,\, \hat D_n^{\text{ref}},\,
  h_{\mathrm{bal},n}^k,\, h_{\mathrm{prev},n}^k,\, h_{\mathrm{stall},n}^k,\,
  \tilde a_n^{\text{norm}},\, a_n^{\text{norm}},\, b_n^{\text{norm}}\big),\\
  e_{mn}^k &= \big(q_{mn}^k/\kappa_{mn},\; q_{nm}^k/\kappa_{mn},\; c_{nm}^{\text{norm}}\big).
\end{aligned}
\label{eq:features}
\end{equation}
The first three node features locate the current demand, generation, and target demand within their local operating ranges, e.g., $\hat D_n^k := 2\frac{D_n^k-\underline D_n}{\overline D_n-\underline D_n}-1$,  %
which lies in $[-1,1]$ within the operating range and is clipped to $[-2,2]$ outside it, since the auxiliary iterate need not be feasible; $h_{\mathrm{bal},n}^k := \frac{D_n^k-G_n^k-\Delta G_n-Q_n^k}{\overline D_n-\underline D_n}$ measures how far the iterate is from local energy balance, and $h_{\mathrm{prev},n}^k$ and $h_{\mathrm{stall},n}^k$, the normalized displacement of the local iterate and its exponentially weighted moving average, capture the progress and persistent slowdowns of the iteration; the remaining node features, and the last edge feature, are cost coefficients normalized by their parameter ranges, fixed a priori (Table~\ref{tab:params}), while the first two edge features express each trade relative to the link capacity. Since both features follow the same template at every node and edge, their dimensions, $d_v = 9$ and $d_e = 3$ are independent of the market size and node degrees.

A shared multilayer perceptron (MLP)~\citep{taud2017multilayer} $\phi_{\mathrm{msg}}$ maps each edge feature to a message in $\mathbb{R}^{d_h}$. The messages are aggregated and passed, along with $h_n^k$, to a Gated Recurrent Unit (GRU)~\citep{cho2014learning}, which updates a private memory $\mathrm{mem}_n^k\in\mathbb{R}^{d_h}$:
\begin{equation}
  h_n^{\mathrm{agg},k} = \frac{1}{\sqrt{|\Omega_n|}}\sum_{m\in\Omega_n}\phi_{\mathrm{msg}}\big(e_{mn}^k\big),
  \qquad
  \mathrm{mem}_n^k = \operatorname{GRU}\big(\operatorname{col}(h_n^k, h_n^{\mathrm{agg},k}),\, \mathrm{mem}_n^{k-1}\big),
  \label{eq:gnn}
\end{equation}
with $\mathrm{mem}_n^{-1}=0$, where $|\Omega_n|\geq1$ since $\mathcal G$ is connected and $N\geq 2$. The symmetric aggregation is invariant to neighbor ordering and yields a fixed-dimensional input independently of $|\Omega_n|$, while the memory remains local and is never communicated. Finally, a node head for demand and generation and a shared edge head for bilateral trades, both with bounded $\tanh$ output activations rescaled to the local operating ranges, produce the local correction $u_n^k\in\mathbb{R}^{2+|\Omega_n|}$. 

We denote the resulting shared local map by $(u_n^k,\mathrm{mem}_n^k)=\Phi_w\big(h_n^k,\{e_{mn}^k\}_{m\in\Omega_n},\mathrm{mem}_n^{k-1}\big)$, with trainable parameters $w$. In particular, we note that, since all trainable maps operate on fixed-dimensional representations and their parameters are shared across the graph, the same parametrization can be deployed on instances with different sizes and trading topologies. 
As detailed in Appendix~\ref{app:gnn1}, the parameters $w$ are learned once, offline, on historical market instances, by minimizing a finite-horizon proxy of~\eqref{eq:ideal_loss} that requires no equilibrium solutions.

\section{Convergence analysis}
\label{sec:guarantees}

In this section, we establish the theoretical guarantees of the scheme introduced in Section~\ref{sec:learned_dr}. All the corresponding proofs are deferred to Appendix~\ref{app:proofs}.
We begin with a feasibility condition that only restricts the market parameters $\theta$, requiring compatibility between the local operational constraints of the prosumers and the reciprocity constraints.

\begin{assumption}[Market feasibility]\label{ass:game}
For every market instance $\eta=(\mathcal{G}, \theta)$ under consideration, the joint feasible set $\mathcal{X}(\theta) = \mathcal{X}^{\text{loc}}(\theta) \cap \mathcal{X}^{rec}$ is nonempty.
\end{assumption}

Under Assumption~\ref{ass:game}, $\mathcal X^{\mathrm{\text{loc}}}(\theta)$ is a compact convex polyhedron and $\mathcal X^{\mathrm{rec}}$ is a closed convex polyhedral set, hence their intersection $\mathcal X(\theta)$ is nonempty, compact, and convex. Moreover, since the local objectives are continuously differentiable, convex, and separable across prosumers, the pseudo-gradient $F(\cdot;\theta)$,  introduced in Definition~\ref{def:vgne}, is monotone.

\begin{lemma}[Existence]\label{lem:existence} Under Assumption~\ref{ass:game}, the P2P market game~\eqref{eq:game} admits at least one v-GNE.
\end{lemma}

Lemma~\ref{lem:existence} shows that the set $\mathrm{SOL}\big(\mathcal{X}(\theta), F(\cdot;\theta)\big)$, while not necessarily a singleton, is nonempty. The remaining question is whether Algorithm~\ref{alg:learned_dr} converges for every choice of trainable parameters. The key observation is that the bounded GNN output, together with the geometric decay $\rho^k$, makes the learned correction summable, i.e., $\{v^k\}_{k\geq0}\in \ell_1$, for any $w$. Algorithm~\ref{alg:learned_dr} can therefore be viewed as a summably perturbed DR iteration, placing it within the class of inexact fixed-point schemes for which convergence of the nominal iteration is preserved~\citep{martin_learning_2026}.
\frenchspacing

\begin{theorem}[Convergence]
\label{thm:convergence}
    Let Assumption~\ref{ass:game} hold, let $\gamma>0$, $\lambda\in(0,2)$, and $\rho\in(0,1)$. Then, for any market instance $\eta$, for every initialization $x^0\in\mathbb{R}^{n_x}$ and every choice of the trainable parameters $w$, there exists $x^\star\in \mathrm{SOL}\big(\mathcal{X}(\theta),   F(\cdot;\theta)\big)$ such that the shadow iterates $\tilde x^k := \operatorname{proj}_{\mathcal{X}^{\mathrm{rec}}}(x^k)$ of Algorithm~\ref{alg:learned_dr} satisfy $\tilde x^k \to x^\star$, i.e., they converge to a v-GNE of the P2P market game~\eqref{eq:game}.
\end{theorem}

The result also makes explicit how the two components of the proposed scheme contribute to its behavior: convergence is inherited from the nominal DR iteration, while the learned correction acts on the transient through a summable perturbation. 

\section{Numerical Results}
\label{sec:numerical}

We evaluate whether a correction trained on a small set of historical market instances reduces the number of iterations Algorithm~\ref{alg:learned_dr} requires to clear new, unseen instances, compared with the baseline DR iteration, and with alternative acceleration strategies.

\subsection{Experimental setup}
\label{sec:setup}

We consider P2P markets with $N = 10$ prosumers, generated according to the model of Section~\ref{sec:probformulation} from ~\citet{cadre_peer-peer_2020}. 
Since historical records of P2P markets are difficult to obtain, we generate synthetic markets. 
For each density $\delta \in \{0.2, 0.4, 0.5, 0.6, 0.8, 1\}$, we draw a base market, i.e., an instance $\eta = (\mathcal G, \theta)$ with $\theta$ drawn from the distributions in Appendix~\ref{app:data} and $\mathcal G$ a random spanning tree to which a fraction $\delta$ of the remaining edges is added uniformly at random, so that $\delta = 1$ yields the complete graph. To emulate historical market data, where recurring instances share the same trading graph and cost structure but differ in operating conditions, we then generate $M = 50$ training, 10 validation, and 30 test instances by rescaling the demand- and supply-related components of $\theta$, namely $(D^{\mathrm{ref}}_n, \underline{D}_n, \overline{D}_n, \underline{G}_n, \overline{G}_n, \Delta G_n)_{n\in\mathcal N}$, by up to $\pm 10\%$.
The DR parameters are tuned by grid search on the training instances and kept identical for all methods, so that any performance difference can be attributed to the acceleration strategy.

We train a separate GNN for each base market by unrolling Algorithm~\ref{alg:learned_dr} over a finite horizon $K$ shorter than the number of DR iterations, and choose $\rho$ so that the correction is mostly switched off by the end of this horizon
(details in Appendix~\ref{app:impl}). All methods run until the fixed-point residual, computed on each method's own iterates (including $v^k$ for Algorithm~\ref{alg:learned_dr}), falls below $\tau = 10^{-5}$, or for at most 4000 iterations. Unless otherwise stated, they start from the same random initial point.

For each test instance $i$, we measure the relative iteration reduction $\Delta_i := (k^i_{\mathrm{DR}} - k^i)/k^i_{\mathrm{DR}}$, where $k^i_{\mathrm{DR}}$ and $k^i$ are the iteration counts of DR and of the evaluated method. 
The main text focuses on three representative densities, $\delta \in \{0.2, 0.6, 1\}$, ranging from the sparsest to the complete trading graph, while Appendix~\ref{app:results} covers all densities.

\subsection{In-distribution performance}

We compare Algorithm~\ref{alg:learned_dr} (DR+GNN) with both learning-based and classical acceleration strategies. We consider a learned warm start (L2WS)~\citep{sambharya2024learning}, which predicts the initial iterate from the market parameters, replacing the random initialization, and is trained, as Algorithm~\ref{alg:learned_dr}, with the self-supervised loss~\eqref{eq:training-loss}, so that neither method requires equilibrium solutions. We include its original centralized version, parametrized by an MLP over all market parameters, and a distributed version, parametrized by a GNN with $L$ message-passing layers, whose rounds are added to its iteration count, with $L = 3$ selected as a trade-off between the information available and the additional communication rounds. As a classical acceleration scheme for fixed-point iterations, we consider Anderson-accelerated DR (A2DR), safeguarded as in~\citet{fu2020anderson} and applied to the same DR operator, whose iterations additionally require global inner products among all prosumers. The centralized L2WS and A2DR thus require global coordination.

\begin{figure}[!ht]
    \centering
    \includegraphics[width=\linewidth]{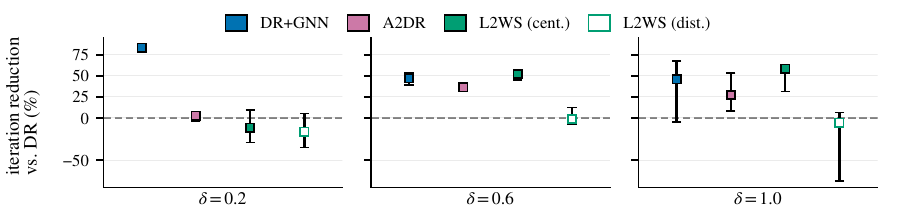}
    \vspace{-2.5em}
    \caption{Iteration reduction w.r.t.\ DR on in-distribution instances
    (median and interquartile range).}
    \label{fig:methods3}
\end{figure}

Fig.~\ref{fig:methods3} shows that Algorithm~\ref{alg:learned_dr} substantially reduces the number of iterations required by DR, as well as their variability across instances (Table~\ref{tab:methods} in Appendix~\ref{app:results}). In contrast, the distributed L2WS provides little or no acceleration: a warm start predicts the solution once, from the information reaching each prosumer within $L$ rounds, whereas the learned correction observes the iterates and only needs to steer an already convergent iteration. Algorithm~\ref{alg:learned_dr} also achieves a larger median reduction than A2DR at every density, and than the centralized L2WS, which exploits the parameters of all prosumers, only at the sparsest density. On the complete graph, however, the reduction is less uniform: the iteration count of Algorithm~\ref{alg:learned_dr} varies little, whereas that of DR varies widely, and some instances that DR already clears quickly are not accelerated.

\begin{figure}[!ht]
    \centering
    \vspace{-1em}
    \includegraphics[width=\linewidth]{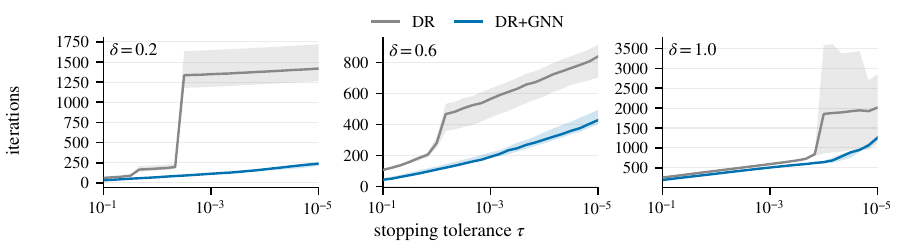}
    \vspace{-2.5em}
    \caption{Iterations to reach a tolerance on the fixed-point
residual (median and interquartile range).}
    \label{fig:tolerance_3}
\end{figure}

Fig.~\ref{fig:tolerance_3} reports the iterations required to reach a given tolerance on the fixed-point residual, and shows that, as the tolerance decreases, DR enters a slow phase, in which its iteration count increases dramatically. In contrast, Algorithm~\ref{alg:learned_dr} avoids this phase, and its iteration count instead grows smoothly. The learned correction therefore does not speed up all iterations uniformly, its benefit is largest when the market must be cleared accurately.

\subsection{Out-of-distribution performance}
We next evaluate, without retraining, two practically motivated out-of-distribution (OOD) shifts that may naturally arise across repeated market-clearing intervals. First, operating conditions may vary beyond the historical data, e.g., across seasons: we increase the level perturbation from $\pm 10\%$ to
$\pm 20\%$, $\pm 30\%$, and $\pm 40\%$. Second, prosumers may temporarily withdraw, e.g., due to maintenance or communication outages, or join after installing new generation assets: we add or remove one prosumer with equal probability, connected to as many random neighbors as the average degree of the base network. The latter test is enabled by the size-independent parametrization of Section~\ref{sec:gnn}.

\begin{figure}[!ht]
    \centering
    \vspace{-1em}
    \includegraphics[width=\linewidth]{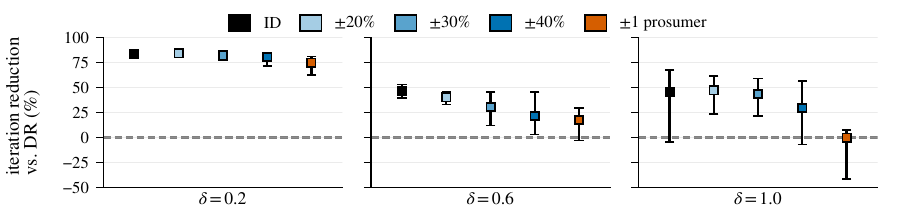}
    \vspace{-2.5em}
    \caption{Iteration reduction of Algorithm~\ref{alg:learned_dr} w.r.t.\ DR under
level shifts and changes of one prosumer, without retraining (median and
interquartile range).}
    \label{fig:ood3}
    \vspace{-2em}

\end{figure}

As shown in Fig.~\ref{fig:ood3}, the acceleration degrades gracefully under level shifts, and remains substantial up to $\pm 40\%$, i.e., four times the variability seen in training. Structural changes are more challenging, since they alter several local subproblems at once and produce neighborhoods never observed during training: the median reduction remains non-negative at every density, but on denser networks a few instances are slowed down, partly because the modified markets are easier for DR and leave less room for acceleration (Table~\ref{tab:ood} in Appendix~\ref{app:results}). In all cases, convergence is preserved by Theorem~\ref{thm:convergence}, so that a mismatch between training and deployment can only affect the speed of clearing.


\section{Conclusion}
\label{sec:conclusion}
We proposed a learning-enhanced market-clearing scheme that exploits historical market data to accelerate the distributed computation of variational equilibria in P2P energy markets. The scheme augments a DR iteration with a GNN-based correction that preserves distributed implementation and convergence to a v-GNE for any trained parameters; since the analysis only requires the baseline to iterate a nonexpansive operator, the construction extends to other market-clearing algorithms. On unseen instances, the correction substantially reduces the number of iterations, and their variability, achieving larger median reductions than a distributed learned warm start and Anderson acceleration, while a centralized warm start remains faster on denser markets. Its gains degrade gracefully under shifts in operating conditions, whereas structural changes to the market remain more challenging.

Future work includes distributed training, to preserve the privacy of the prosumers' historical data throughout the learning process, and generalization guarantees, to make the accelerated clearing time certifiable. Finally, the approach should be validated on real market data and on larger communities, to enable P2P energy markets at scale.

\subsection*{Generative AI Use}
Generative AI tools were used to assist with coding and with writing the documentation of the code. 
The authors reviewed and verified all AI-assisted content and take responsibility for the final content of the paper.

\acks{This work was partially supported by Digital Futures and by the Wallenberg AI, Autonomous Systems and Software Program (WASP) funded by the Knut and Alice Wallenberg Foundation.}

\bibliography{l4dc2027-sample}

\newpage
\appendix
\counterwithin{table}{section}
\newif\ifshortbc
\shortbctrue   %

\newcommand{\bc}[1]{%
  \ifshortbc (#1)\else\citep[#1]{bauschke2017convex}\fi}

\newcommand{\bcnote}{%
  \ifshortbc Unless otherwise stated, numbered results refer to \citet{bauschke2017convex}.\fi}

\section{Technical Proofs}
\label{app:proofs}
For notational simplicity, we fix a market instance $\eta=(\mathcal G,\theta) $ and omit the explicit dependence on $\theta$, writing $F(x):=F(x;\theta)$, $\mathcal X^{\mathrm{\text{loc}}}:=\mathcal X^{\mathrm{\text{loc}}}(\theta)$, and $\mathcal X:=\mathcal X(\theta)$. \bcnote

\subsection{Proof of Lemma~\ref{lem:existence}}
\label{app:proofLemma}

Since each $J_n$ is convex and continuously differentiable in $x_n$, and depends only on $x_n$, the pseudo-gradient $F=\operatorname{col}\big((\nabla_{x_n}J_n)_{n\in\mathcal N}\big)$ is monotone and continuous, hence maximally monotone~\bc{Cor.~20.28}. As $\mathcal X$ is nonempty, closed, and convex, $N_{\mathcal X}$ is maximally monotone~\bc{Ex.~20.26}, and so is $T:=F+N_{\mathcal X}$, since $\operatorname{dom}F=\mathbb R^{n_x}$~\bc{Cor.~25.5(i)}. Because $\operatorname{dom}T=\mathcal X$ is bounded, $T$ is surjective~\bc{Cor.~21.25}. Hence, there exists $x^\star\in\mathcal X$ such that $0\in F(x^\star)+N_{\mathcal X}(x^\star)$, i.e., $x^\star\in\operatorname{SOL}(\mathcal X,F)$, and therefore $x^\star$ is a v-GNE by Definition~\ref{def:vgne}.

\subsection{Proof of Theorem~\ref{thm:convergence}}
\label{app:proof}
We first identify the baseline component of Algorithm~\ref{alg:learned_dr} with a relaxed DR iteration. Define the two operators 
\begin{equation}\label{eq:AB}
  \mathcal{A} := F + N_{\mathcal{X}^{\mathrm{loc}}},
  \qquad
  \mathcal{B} := N_{\mathcal{X}^{\mathrm{rec}}},
\end{equation}
which isolate the local operational constraints and the coupling reciprocity constraints, respectively. Let $J:=\sum_{n\in\mathcal N}J_n$. By separability and differentiability of the local costs, $F=\nabla J=\partial J$~\bc{Prop.~17.31(i)}. Since $\operatorname{dom}J=\mathbb R^{n_x}$, the sum rule gives $\mathcal A=\partial\big(J+\iota_{\mathcal X^{\mathrm{loc}}}\big)$~\bc{Cor.~16.48(iii)}, where $\iota_C$ denotes the indicator function of a set $C$. Hence, $\mathcal A$ is maximally monotone~\bc{Thm.~20.25}, and its resolvent $J_{\gamma\mathcal A}=\operatorname{prox}_{\gamma(J+\iota_{\mathcal X^{\mathrm{loc}}})}$~\bc{Ex.~23.3} decouples, by separability, into the node-wise minimizations of line~6~\bc{Prop.~24.11 \& Def.~12.23}. Similarly, $\mathcal B$ is maximally monotone~\bc{Ex.~20.26}, and, since $\gamma N_{\mathcal X^{\mathrm{rec}}}=N_{\mathcal X^{\mathrm{rec}}}$, $J_{\gamma\mathcal B}=\operatorname{proj}_{\mathcal X^{\mathrm{rec}}}$~\bc{Ex.~23.4}. Since each constraint in~\eqref{eq:reciprocity} involves a distinct pair of trade variables, this projection is computed edge-wise in lines~4, yielding $\tilde x^k=J_{\gamma\mathcal B}(x^k)$, while line~5 evaluates the reflected resolvent $R_{\gamma\mathcal B}:=2J_{\gamma\mathcal B}-\mathrm{Id}$, as $r^k=2\tilde x^k-x^k$. Hence, line~6 evaluates $z^k=J_{\gamma\mathcal A}R_{\gamma\mathcal B}(x^k)$, where both resolvents are computed from local data and one-hop communication only. Defining the DR operator $T_{\mathrm{DR}}:=J_{\gamma\mathcal A}R_{\gamma\mathcal B}+\mathrm{Id}-J_{\gamma\mathcal B}$, so that $T_{\mathrm{DR}}(x^k)-x^k=z^k-\tilde x^k$, line~7 of Algorithm~\ref{alg:learned_dr} can be written at the aggregate level as
\begin{equation}
    \hat x^{k}
    =
    x^k
    +
    \lambda
    \big(
        T_{\mathrm{DR}}(x^k)-x^k
    \big),
    \label{eq:baseline-dr-update}
\end{equation}
which is precisely the relaxed DR recursion associated with~\eqref{eq:AB}~\bc{Thm.~26.11}.

We next identify the zeros of $\mathcal A+\mathcal B$ with the v-GNEs of the market game. Since $\mathcal X^{\mathrm{loc}}$ and $\mathcal X^{\mathrm{rec}}$ are polyhedral and $\mathcal X=\mathcal X^{\mathrm{loc}}\cap\mathcal X^{\mathrm{rec}}\neq\varnothing$ by Assumption~\ref{ass:game}, the polyhedral subdifferential sum rule~\bc{Thm.~16.47(iii)} gives $N_{\mathcal X^{\mathrm{loc}}}+N_{\mathcal X^{\mathrm{rec}}}=\partial(\iota_{\mathcal X^{\mathrm{loc}}}+\iota_{\mathcal X^{\mathrm{rec}}})=\partial\iota_{\mathcal X}=N_{\mathcal X}$. Hence, $\mathcal A+\mathcal B=F+N_{\mathcal X}$ and\looseness=-1
\begin{equation}
    \operatorname{zer}(\mathcal A+\mathcal B)
    =
    \left\{
        x\in\mathcal X
        \mid
        -F(x)\in N_{\mathcal X}(x)
    \right\}
    =
    \operatorname{SOL}(\mathcal X,F),
    \label{eq:zero-vgne}
\end{equation}
which is nonempty by Lemma~\ref{lem:existence}. Since $\mathcal A$ and $\mathcal B$ are maximally monotone, their reflected resolvents $R_{\gamma\mathcal A}$ and $R_{\gamma\mathcal B}$ are nonexpansive~\bc{Cor.~23.11}. 
As $T_{\mathrm{DR}}=\tfrac12\big(\mathrm{Id}+R\big)$, where $R:=R_{\gamma\mathcal A}R_{\gamma\mathcal B}$, it follows that $T_{\mathrm{DR}}$ is firmly nonexpansive, with $\operatorname{Fix}T_{\mathrm{DR}}=\operatorname{Fix}(R_{\gamma\mathcal A}R_{\gamma\mathcal B})$, and
\begin{equation}
    \operatorname{zer}(\mathcal A+\mathcal B)
    =
    J_{\gamma\mathcal B}
    \left(
        \operatorname{Fix}T_{\mathrm{DR}}
    \right)
    \label{eq:shadow-characterization}
\end{equation}
~\bc{Prop.~26.1(iii)}. Since $\operatorname{zer}(\mathcal A+\mathcal B)\neq\varnothing$, it follows that $\operatorname{Fix}T_{\mathrm{DR}}\neq\varnothing$.

We now account for the learned component. Line~8 of Algorithm~\ref{alg:learned_dr} evaluates the local correction $v_n^k$, and line~9 adds it to $\hat x_n^k$. 
Defining $v^k:=\operatorname{col}\big((v_n^k)_{n\in\mathcal N}\big)$ and
$u^k := \operatorname{col}\big((u_n^k)_{n\in\mathcal N}\big)$, by construction of $\Phi_w$, the $\tanh$ output activations, rescaled by fixed local scales, bound every component of $u^k$, so that there exists $C(\eta)>0$ such that $\|u^k\|\le C(\eta)$ for all $k\ge0$, independently of $w$. Hence,
by~\eqref{eq:decaying_perturbation}, $\|v^k\|=\rho^k\|u^k\|\le C(\eta)\,\rho^k$,
and, since $\rho\in(0,1)$,
$\sum_{k=0}^{\infty}\|v^k\|\le C(\eta)/(1-\rho)<\infty$, for any choice of
the trainable parameters $w$, i.e., $\{v^k\}_{k\ge0}\in\ell_1$.

By~\eqref{eq:baseline-dr-update} and line~9, the iteration reads
\begin{equation}\label{eq:perturbed-dr}
  x^{k+1} = x^k + \lambda\big(T_{\mathrm{DR}}(x^k) - x^k\big) + v^k
          = x^k + \tfrac{\lambda}{2}\big(R(x^k) - x^k\big) + v^k,
\end{equation}
since $T_{\mathrm{DR}} = \tfrac12(\mathrm{Id} + R)$.
The learned map therefore leaves the DR operator $T_{\mathrm{DR}}$ unchanged and enters only as an additive perturbation of its auxiliary fixed-point sequence. 

As $\tfrac{\lambda}{2}\in(0,1)$, this is an inexact Krasnosel'ski\u{\i}--Mann (KM) iteration of the nonexpansive map $R$, whose fixed-point set is nonempty, with summable error sequence $\{v^k\}_{k\geq0}$, whose bound holds uniformly along any trajectory.

By the convergence of summably perturbed KM iterations~\citep[Prop.~3]{martin_learning_2026}, $x^k$ converges to some $x_{\mathrm{aux}}^\star\in\operatorname{Fix}T_{\mathrm{DR}}$. Finally, $J_{\gamma\mathcal B}$ is firmly nonexpansive, hence continuous, so that
$
    \tilde x^k
    =
    J_{\gamma\mathcal B}(x^k)
    \to
    J_{\gamma\mathcal B}(x_{\mathrm{aux}}^\star)
    =:
    x^\star,
$
and $x^\star\in\operatorname{zer}(\mathcal A+\mathcal B)=\operatorname{SOL}(\mathcal X,F)$ by~\eqref{eq:shadow-characterization} and~\eqref{eq:zero-vgne}. Therefore, for every initialization and every choice of the trainable parameters, the shadow iterates of Algorithm~\ref{alg:learned_dr} converge to a v-GNE of the P2P market game.

 \section{Offline training }
\label{app:gnn1}
The parameters $w$ are learned once, offline and centrally, from historical market data, i.e., a dataset of $M$ past market-clearing instances $\mathcal D_{\mathrm{tr}}=\{\eta^{i}\}_{i=1}^{M}$, viewed as samples from $\mathbb P$. For each one, Algorithm~\ref{alg:learned_dr} and the baseline DR method are unrolled for $K$ iterations from the same initial point $x^{0,i}$, yielding $\{x^{k,i}(w)\}_{k=0}^{K}$ and $\{x_{\mathrm{DR}}^{k,i}\}_{k=0}^{K}$, respectively, the latter is computed without gradient tracking and serves only as an offline reference. Since the distance-based criterion in~\eqref{eq:ideal_loss} requires the unknown equilibrium set, we measure finite-time progress through the normalized iterate residuals
$
\hat r_i^k(w)
=
\|x^{k,i}(w)-x^{k-1,i}(w)\|/(\|\sigma_i\|+\varepsilon),
$
and analogously $\hat r_{\mathrm{DR},i}^k$, where $\sigma_i$ collects the half-widths of the demand and generation intervals and the link capacities of $\eta^i$ and $\varepsilon>0$ is a regularization constant. The finite-horizon training loss for sample $i$ is
\begin{equation}
\mathcal{L}_K^i(w)
=
\frac{1}{K}\sum_{k=1}^{K}
\mu^i_k\,                                                      
\ell\big(\hat r_i^k(w)-\hat r_{\mathrm{DR},i}^k\big)
+
\frac{\beta}{K-1}\sum_{k=1}^{K-1}
\big[\hat r_i^{k+1}(w)-\hat r_i^k(w)\big]_+,
\label{eq:training-loss}
\end{equation}
with $\ell(a)=\alpha_{\mathrm{behind}}[a]_+-\alpha_{\mathrm{ahead}}[-a]_+$, $[a]_+:=\max\{a,0\}$, $\alpha_{\mathrm{behind}}>\alpha_{\mathrm{ahead}}>0$, and $\beta\geq0$. The first term rewards progress ahead of the DR baseline while penalizing lagging behind it, the unequal slopes ensure that the baseline residual shapes the gradient with respect to $w$, rather than acting as an additive constant. The second term penalizes residual increases, promoting rapid and regular decay. Neither term requires equilibrium labels or precomputed solutions, hence, training only requires the past market instances, not their cleared outcomes. 
The weights $\mu^i_k>0$, which play the role of $\alpha_k$ in~\eqref{eq:ideal_loss}, rebalance the loss over the horizon. Since the residuals decay by several orders of magnitude, uniform weights would let the first iterations dominate, whereas the slow phase of DR, which determines its iteration count, would contribute little. The weights are computed from the DR rollout only and normalized to unit mean, and either grow with $k$, or are adapted to the DR trajectory, with larger weights at the iterations where the DR residual decreases slowly.

Supervised surrogates of~\eqref{eq:ideal_loss}, e.g., regressing the iterates onto an equilibrium of each training instance, can also be used, alone or combined with~\eqref{eq:training-loss}, without affecting the convergence guarantees, but require the solution to be stored with the historical data or computed offline, or, since the equilibrium need not be unique, the optimal aggregate cost $J^{\star,i}$ to be used instead.

Since $\mathbb P$ is unknown, $w^\star \in \arg\min_w \frac{1}{M}\sum_{i=1}^{M}
\mathbb E_{x^{0,i}}\big[\mathcal L_K^i(w)\big]$, where $x^{0,i}$ is drawn uniformly within the local bounds and resampled at every epoch, is approximately computed by backpropagating through the learned $K$-step unroll.
The distributed and convergence requirements in~\eqref{eq:optimal-iteration-design-problem} are enforced by the baseline iteration and the correction architecture, while learning only improves the finite-horizon objective.  
Once trained, $w^\star$ is shared with all prosumers, which evaluate $\Phi_{w^\star}$ using only local information and their private memory, neither the baseline rollout nor any central coordination is required during market operation.

 \section{Implementation details}
 \label{app:impl-details}
 \subsection{Market data}
\label{app:data}

The parameters of each base market are drawn independently for each prosumer or trading link from the distributions in Table~\ref{tab:params}.
As in \citet{cadre_peer-peer_2020}, trading preferences are heterogeneous and asymmetric, and the curvatures of the generation cost and of the usage comfort are of the same order. The demand interval of each prosumer
is centered at its local supply $D^0_n$, at an interior generation level, so that $(D_n, G_n, q_n) = (D^0_n, G^0_n, 0)$ is feasible and Assumption~\ref{ass:game} holds by construction. \looseness-1

Each instance is obtained from its base market through factors common to all prosumers. The levels of demand, flexible generation, and renewable generation are rescaled by factors drawn from $U(0.9, 1.1)$, and the widths of the operating intervals by factors drawn from $U(0.96, 1.04)$. In the out-of-distribution tests, the level factors are drawn from $U(0.8, 1.2)$, $U(0.7, 1.3)$, and $U(0.6, 1.4)$, respectively, and the width factors from $U(0.92, 1.08)$. In the test with one prosumer more or less, the removed prosumer is chosen uniformly at random, while the added one has parameters drawn from the distributions in Table~\ref{tab:params}.
Instances with a disconnected trading graph or violating Assumption~\ref{ass:game} are discarded and resampled.

\begin{table}[h]
\centering
\label{tab:params}
\small
\begin{tabular}{ll}
\toprule
Parameter & Distribution \\
\midrule
$\Delta G_n$ & $U(0.5, 2)$ \\
$\underline{G}_n$, $\overline{G}_n$ & $U(0, 1)$, $U(2, 8)$ \\
$[\underline{D}_n, \overline{D}_n]$ & centered at local supply, half-width $U(0.4, 1)$ \\
$a_n$, $\tilde a_n$, $b_n$ & $U(0.5, 2)$, $U(0.5, 1.5)$, $U(0.2, 1)$ \\
$c_{nm}$ & $U(0.05, 0.18)$\\
$\kappa_{mn} = \kappa_{nm}$ & $U(1, 4)$ \\
\bottomrule
\end{tabular}
\caption{Distribution of the parameters of the base markets (powers in kW).}
\end{table}

 \subsection{Training and computation}
\label{app:impl}
All methods are implemented in Python 3.10 and PyTorch 2.11, in single precision, and
run on a single core of an Intel Xeon w7-3455 CPU (24 cores, 48 threads, 250\,GB RAM). The local proximal step is solved by bisection on the multiplier of the local energy balance, with $30$ bisection steps. 
For each base market, $\Phi_w$ uses hidden dimension $d_h=32$ and one hidden layer per MLP, and is trained for $25$ epochs using Adam with learning rate $3 \cdot 10^{-3}$. The training horizon $K$ is set to half the average number of DR iterations on 40 separate calibration instances of the same base market, and $\rho = 1 - 10/K$, so that $\rho^K \approx e^{-10}$ and the correction is mostly switched off by the end of the training horizon.
The DR parameters are tuned for each base market by grid search on the training instances, 
all other hyperparameters are selected on the validation instances. Among these, the slopes of the loss, the type of iteration weights, and the amplitude of the correction are selected for each density, while the remaining ones, including the number of message-passing layers $L=3$ of the distributed L2WS and the
memory and regularization of A2DR, are shared across densities.

At every iteration, each prosumer solves its local proximal step (line~6 of Algorithm~\ref{alg:learned_dr}), which dominates the computational cost of an iteration of DR, and, in Algorithm~\ref{alg:learned_dr}, additionally evaluates the learned correction $\Phi_w$ (line~8), whose cost scales with the number of trading partners, since $\Phi_w$ processes one message per neighbor, i.e., has complexity $O(|\Omega_n|)$.
Table~\ref{tab:timing} reports both times and their ratio, which we refer to as overhead and which approximates the additional cost of an iteration of Algorithm~\ref{alg:learned_dr} relative to one of DR. This overhead, between $47\%$ and $97\%$, is compensated by the reduction in the number of iterations, so that the total computation time decreases at all densities except $\delta=1$,
where it is comparable to that of DR. In a P2P market, moreover, each iteration also entails a round of communication, so that the number of rounds remains the relevant measure of clearing cost.

\begin{table}[H]
\centering
\label{tab:timing}
\small
\setlength{\tabcolsep}{4pt}
\begin{tabular}{lcccc}
\toprule
$\delta$ & Local prox. [ms/agent/iter.] & $\Phi_w$ [ms/agent/iter.] & Overhead & Training [min] \\
\midrule
$0.2$ & 0.760 & 0.359 & 47\% & 47.6 \\
$0.4$ & 0.742 & 0.578 & 78\% & 38.2 \\
$0.5$ & 0.789 & 0.461 & 58\% & 67.1 \\
$0.6$ & 0.760 & 0.471 & 62\% & 35.5 \\
$0.8$ & 0.771 & 0.720 & 93\% & 143.6 \\
$1.0$ & 0.811 & 0.788 & 97\% & 169.8 \\
\bottomrule
\end{tabular}
\caption{Computation time per prosumer and iteration of the local proximal step and of $\Phi_w$, overhead, i.e., the ratio between the two, and training time per base market.}
\end{table}

\section{Detailed numerical results}
\label{app:results}

 \subsection{In-distribution test}
 \label{app:idtable}

\begin{figure}[!ht]
    \centering
    \includegraphics[width=\linewidth]{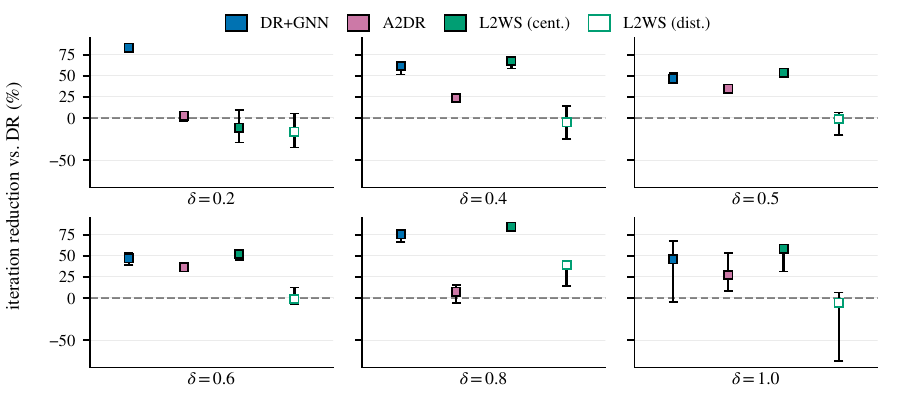}
    \vspace{-2.5em}
    \caption{Iteration reduction w.r.t.\ DR on in-distribution instances, for all densities (median and interquartile range).}
    \label{fig:methods}
    \vspace{-1.5em}
\end{figure}

\begin{figure}[!ht]
    \centering
    \includegraphics[width=\linewidth]{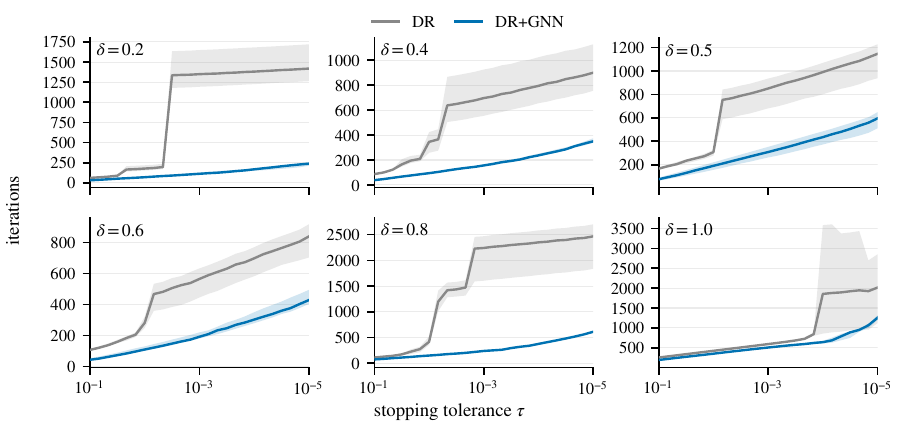}
    \vspace{-2.5em}
    \caption{Iterations to reach a tolerance on the fixed-point
residual (median and interquartile range).}
    \label{fig:tolerance_all}
\end{figure}

Fig.~\ref{fig:methods} reports the relative iteration reduction for all densities, and Table~\ref{tab:methods} the corresponding iteration counts, with the fastest distributed method in bold, while Fig.~\ref{fig:tolerance_all} reports the iterations required to reach each tolerance on the fixed-point residual. The learned correction reduces the dispersion across instances by a factor of up to about $70$, so that the number of rounds required to clear the market becomes considerably more predictable.

\begin{table}[H]
\centering
\label{tab:methods}
\small
\setlength{\tabcolsep}{4pt}
\begin{tabular}{lcccccc}
\toprule
Method & $\delta=0.2$ & $\delta=0.4$ &$\delta=0.5$ & $\delta=0.6$ &$\delta=0.8$ & $\delta=1.0$ \\
\midrule
DR (dist.)
& $1419 \pm 229$& $900 \pm 186$& $1146 \pm 145$& $838 \pm 107$& $2466 \pm 434$& $2274 \pm 1364$ \\
\midrule
DR+GNN (dist.)
& $\mathbf{238 \pm 23}$& $\mathbf{350 \pm 17}$& $\mathbf{596 \pm 69}$& $\mathbf{428 \pm 46}$& $\mathbf{610 \pm 6}$& $\mathbf{1252 \pm 59}$ \\

L2WS (dist.)
& $1634 \pm 45$& $948 \pm 52$& $1155 \pm 27$& $808 \pm 93$& $1550 \pm 25$& $3526 \pm 786$ \\
\midrule
L2WS (cent.)
& $1560 \pm 54$& $308 \pm 16$& $514 \pm 26$& $387 \pm 52$& $389 \pm 19$& $906 \pm 395$ \\
A2DR
& $1343 \pm 327$& $673 \pm 186$& $714 \pm 107$& $512 \pm 80$& $1997 \pm 593$& $1091 \pm 592$ \\
\bottomrule
\end{tabular}
\caption{Iterations to meet the stopping criterion on in-distribution test instances (median $\pm$ half interquartile range).}
\vspace{-2em}
\end{table}

\subsection{Out-of-distribution test}
\label{app:oodtable}
\begin{figure}[!ht]
    \centering
    \includegraphics[width=\linewidth]{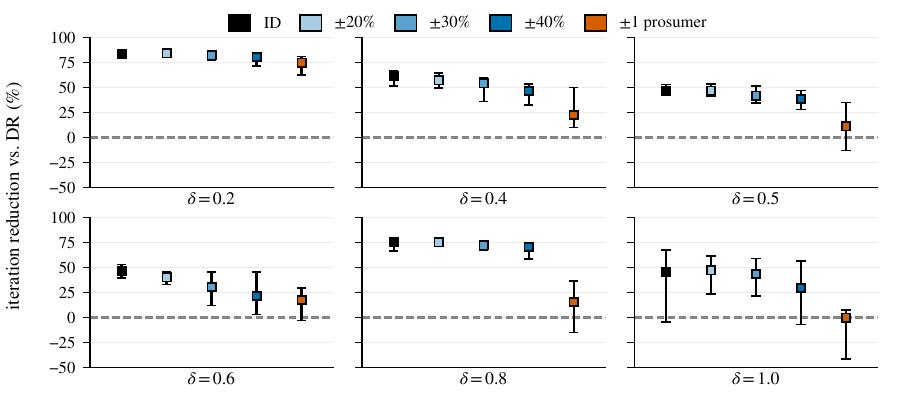}
    \vspace{-2.5em}
    \caption{Iteration reduction of Algorithm~\ref{alg:learned_dr} w.r.t.\ DR under
level shifts and changes of one prosumer, without retraining, for all densities (median and
interquartile range).}
    \label{fig:ood}
\end{figure}

Fig.~\ref{fig:ood} reports the relative iteration reduction under distribution shifts for all densities, and Table~\ref{tab:ood} the corresponding iteration counts. 
 Under level shifts, the median iteration count of DR varies little compared with its spread across instances, so that the decrease in relative reduction is mainly due to the learned correction, whose median iteration count and dispersion grow with the magnitude of the shift, e.g., from $238 \pm 23$ to $324 \pm 123$ at $\delta = 0.2$, while remaining well below those of DR.
 Under the structural shift, the dispersion of Algorithm~\ref{alg:learned_dr} increases markedly at every density, consistently with the presence of neighborhoods never observed during training, whereas its median iteration count remains below that of DR.
 
\newpage
\begin{table}[htbp]
\centering
\label{tab:ood}
\small
\setlength{\tabcolsep}{3pt}
\begin{tabular}{llcccccc}
\toprule
Test set & Method & $\delta=0.2$ & $\delta=0.4$ & $\delta=0.5$ & $\delta=0.6$ & $\delta=0.8$ & $\delta=1.0$ \\
\midrule
\multirow{2}{*}{ID} & DR & $1419\pm229$ & $900\pm186$ & $1146\pm145$ & $838\pm107$ & $2466\pm434$ & $2274\pm1364$ \\
& DR+GNN & $238\pm23$ & $350\pm17$ & $596\pm69$ & $428\pm46$ & $610\pm6$ & $1252\pm59$ \\
\midrule
\multirow{2}{*}{$\pm20\%$} & DR & $1582\pm216$ & $916\pm167$ & $1148\pm120$ & $766\pm117$ & $2473\pm315$ & $2460\pm758$ \\
& DR+GNN & $264\pm28$ & $364\pm51$ & $568\pm64$ & $462\pm39$ & $618\pm9$ & $1322\pm64$ \\
\midrule
\multirow{2}{*}{$\pm30\%$} & DR & $1616\pm290$ & $944\pm128$ & $1141\pm142$ & $786\pm110$ & $2474\pm344$ & $2548\pm751$ \\
& DR+GNN & $285\pm67$ & $397\pm123$ & $624\pm67$ & $492\pm91$ & $647\pm42$ & $1276\pm84$ \\
\midrule
\multirow{2}{*}{$\pm40\%$} & DR & $1660\pm384$ & $904\pm147$ & $1132\pm138$ & $822\pm79$ & $2368\pm356$ & $2392\pm926$ \\
& DR+GNN & $324\pm123$ & $436\pm179$ & $688\pm110$ & $570\pm140$ & $680\pm119$ & $1364\pm204$ \\
\midrule
\multirow{2}{*}{$\pm1\text{ prosumer}$} & DR & $1634\pm572$ & $769\pm193$ & $967\pm174$ & $733\pm258$ & $1028\pm297$ & $1517\pm866$ \\
& DR+GNN & $299\pm100$ & $532\pm131$ & $747\pm241$ & $598\pm256$ & $786\pm197$ & $1392\pm1129$ \\
\bottomrule
\end{tabular}
\caption{In- and out-of-distribution test sets (median $\pm$ half interquartile range).}
\end{table}

\end{document}